\documentclass[preprint,aps,prl,floats,a4paper,superscriptaddress,preprintnumbers,showpacs,tightenlines]{revtex4}
\draft 
\usepackage{amsmath}
\usepackage{graphicx}
\usepackage{epsfig}

\newcommand{\apkg}{B \rightarrow \phi K \gamma}
\newcommand{\pkg}{B^- \rightarrow \phi K^- \gamma}
\newcommand{\pkog}{\bar{B}^0 \rightarrow \phi \bar{K}^0 \gamma}

\begin{document}

\setlength{\unitlength}{1pt}

\epsfysize3cm
\epsfbox{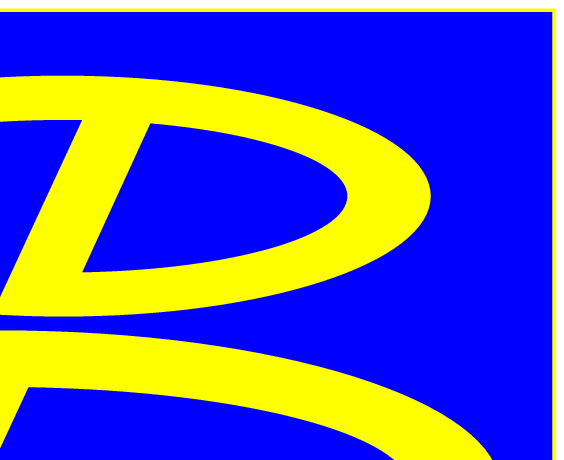}    
\begin{flushright}
\vskip -3cm
\noindent
\hspace*{4.in}Belle Preprint 2003-11 \\
\hspace*{4.in}KEK Preprint 2003-43 \\
\end{flushright}


\

\

\

\title{\large
\bf Observation of radiative {\boldmath{$B$}} $\rightarrow$ {\boldmath{$\phi K \gamma$}} decays.}

\affiliation{Budker Institute of Nuclear Physics, Novosibirsk}
\affiliation{University of Cincinnati, Cincinnati, Ohio 45221}
\affiliation{University of Hawaii, Honolulu, Hawaii 96822}
\affiliation{High Energy Accelerator Research Organization (KEK), Tsukuba}
\affiliation{Hiroshima Institute of Technology, Hiroshima}
\affiliation{Institute of High Energy Physics, Chinese Academy of Sciences, Beijing}
\affiliation{Institute of High Energy Physics, Vienna}
\affiliation{Institute for Theoretical and Experimental Physics, Moscow}
\affiliation{J. Stefan Institute, Ljubljana}
\affiliation{Kanagawa University, Yokohama}
\affiliation{Korea University, Seoul}
\affiliation{Kyungpook National University, Taegu}
\affiliation{Institut de Physique des Hautes \'Energies, Universit\'e de Lausanne, Lausanne}
\affiliation{University of Ljubljana, Ljubljana}
\affiliation{University of Maribor, Maribor}
\affiliation{University of Melbourne, Victoria}
\affiliation{Nagoya University, Nagoya}
\affiliation{Nara Women's University, Nara}
\affiliation{National Lien-Ho Institute of Technology, Miao Li}
\affiliation{Department of Physics, National Taiwan University, Taipei}
\affiliation{H. Niewodniczanski Institute of Nuclear Physics, Krakow}
\affiliation{Nihon Dental College, Niigata}
\affiliation{Niigata University, Niigata}
\affiliation{Osaka City University, Osaka}
\affiliation{Panjab University, Chandigarh}
\affiliation{Peking University, Beijing}
\affiliation{Princeton University, Princeton, New Jersey 08545}
\affiliation{University of Science and Technology of China, Hefei}
\affiliation{Seoul National University, Seoul}
\affiliation{Sungkyunkwan University, Suwon}
\affiliation{University of Sydney, Sydney NSW}
\affiliation{Toho University, Funabashi}
\affiliation{Tohoku Gakuin University, Tagajo}
\affiliation{Tohoku University, Sendai}
\affiliation{Department of Physics, University of Tokyo, Tokyo}
\affiliation{Tokyo Institute of Technology, Tokyo}
\affiliation{Tokyo Metropolitan University, Tokyo}
\affiliation{Tokyo University of Agriculture and Technology, Tokyo}
\affiliation{University of Tsukuba, Tsukuba}
\affiliation{Utkal University, Bhubaneswer}
\affiliation{Virginia Polytechnic Institute and State University, Blacksburg, Virginia 24061}
\affiliation{Yokkaichi University, Yokkaichi}
\affiliation{Yonsei University, Seoul}
  \author{A.~Drutskoy}\affiliation{Institute for Theoretical and Experimental Physics, Moscow} 
  \author{K.~Abe}\affiliation{High Energy Accelerator Research Organization (KEK), Tsukuba} 
  \author{K.~Abe}\affiliation{Tohoku Gakuin University, Tagajo} 
  \author{T.~Abe}\affiliation{High Energy Accelerator Research Organization (KEK), Tsukuba} 
  \author{H.~Aihara}\affiliation{Department of Physics, University of Tokyo, Tokyo} 
  \author{M.~Akatsu}\affiliation{Nagoya University, Nagoya} 
  \author{Y.~Asano}\affiliation{University of Tsukuba, Tsukuba} 
  \author{V.~Aulchenko}\affiliation{Budker Institute of Nuclear Physics, Novosibirsk} 
  \author{T.~Aushev}\affiliation{Institute for Theoretical and Experimental Physics, Moscow} 
  \author{A.~M.~Bakich}\affiliation{University of Sydney, Sydney NSW} 
% ...
\collaboration{The Belle Collaboration}

\date{\today}

\tighten

\begin{abstract}
The radiative decay $\apkg$ is observed for the first time.
The branching fraction for the charged $\pkg$ decay mode is measured to be
\mbox{${\cal B}(\pkg) = (3.4\pm 0.9 \pm 0.4) \times 10^{-6}$}.
The photon energy distribution for the $\pkg$ decay is presented.
The signal for the neutral $\pkog$ decay mode is not 
statistically significant and 
an upper limit, \mbox{${\cal B}(\pkog)\,<\,8.3\,\times\,10^{-6}$} at $90\%\ 
{\rm CL}$, is set.
The analysis is based on a dataset of 90~fb$^{-1}$ collected 
by the Belle experiment at the $e^+ e^-$ asymmetric collider KEKB.
\end{abstract}


\maketitle


Radiative penguin $B$ meson decays provide an important tool
to search for physics beyond the Standard Model. Recent 
experimental studies 
\cite{bellei,bellee,bellekp,cleoi,cleoe,babari,babare,alephi} 
of inclusive and exclusive radiative $B$ decays are 
in good agreement with Standard Model predictions \cite{excl,incl}.
New information on radiative $B$ decays is
important to further test theoretical models.
In this analysis the exclusive decay mode $\apkg$
is studied for the first time.
The experimental techniques used in this analysis are similar to 
those used in the recently published studies of 
\mbox{$B^0 \rightarrow K^+ \pi^- \gamma$} and 
$B^+ \rightarrow K^+ \pi^- \pi^+ \gamma$ decays
by the Belle collaboration \cite{bellekp}.
Measurement of the branching fraction for the $\apkg$ decay
and its contribution to inclusive radiative $B \to X_s \gamma$ decays
is important to constrain theoretical models.
Moreover, due to the narrow width of the $\phi$ resonance, 
exclusive $\apkg$ decays are well-separated from background
and can be effectively used for measurements 
of the photon momentum over a wide interval.
Such measurements 
can shed some light on the behavior of the photon momentum spectrum
in inclusive decays, where the theoretically interesting region 
below 2\,GeV/c is difficult to study experimentally 
because of large backgrounds.
The decay channel $\pkog$ can also be used in future high statistics
measurements of time-dependent $CP$ violation parameters. 
With a larger dataset, such three-body hadronic final state decays could 
also be used for angular distribution measurements \cite{phpol}.
The decay $\pkg$ 
(charge conjugate modes are implied throughout this letter)
can be described by conventional radiative
penguin diagrams with the creation of an additional $s\bar{s}$ pair (for example Fig.~1).
\begin{figure}[h!]
\vspace{-1.5cm}
\begin{center}
\hspace{-0.8cm}
\epsfig{file=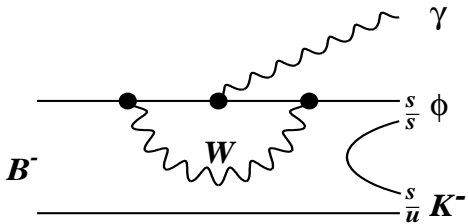,width=7cm,height=7cm}
\end{center}
\vspace{-1.8cm}
\caption{ A penguin diagram for $\pkg$ with $s\bar{s}$ pair 
creation.}
\label{diagramms}
\end{figure}

The 90~fb$^{-1}$ data sample containing $(95.8 \pm 0.5)\times 10^6$
produced $B\bar{B}$ pairs
was collected with the Belle detector \cite{BELLE_DETECTOR} at 
KEKB \cite{aKEKB},
an asymmetric energy double-storage-ring collider with 8\,GeV electrons 
and 3.5\,GeV positrons.
Belle is a general-purpose large-solid-angle detector 
that consists of a three-layer Silicon Vertex Detector (SVD),
a 50-layer Central Drift Chamber (CDC), an array of 
Aerogel \v{C}erenkov Counters (ACC),  a Time-of-Flight
Counter system (TOF), and a CsI(Tl) Electromagnetic 
Calorimeter (ECL) located inside a superconducting
solenoid coil with a 1.5 T magnetic field. An iron
flux-return located outside the coil is instrumented
to identify $K^0_L$ and muons (KLM).
The detector is described in detail elsewhere \cite{BELLE_DETECTOR}.
A detailed GEANT-based simulation of the Belle detector is used
to produce Monte Carlo (MC) event samples and determine efficiencies.

Charged tracks with impact parameters less than 2~cm radially and
less than 5~cm in $z$ (the $z$ axis is antiparallel to the 
positron beam direction)
are used. Kaon and pion mass hypotheses are assigned to charged 
tracks with a momentum larger than 100 MeV/c,
using a likelihood ratio ${\cal L}_{K/\pi}$, obtained 
by combining information from the CDC ($dE/dx$), ACC and TOF systems.
The likelihood ratio ${\cal L}_{K/\pi}$ ranges between 0 and 1 and is 
required to be larger than 0.8 for the kaon 
candidates and less than 0.8 for the pion candidates.
A relaxed likelihood ratio requirement ${\cal L}_{K/\pi}>\,0.4$ 
is applied for the kaon candidates used to reconstruct $\phi$ mesons.

A candidate primary photon ($\gamma$) is required to 
have an energy $E^*_{\gamma}$ in the $\Upsilon$(4S)
center-of-mass (CM) frame 
between 2.0 and 2.7 GeV and to lie within the acceptance
of the barrel ECL ($33^{\circ} < \theta_{\gamma} < 128^{\circ}$).
The main background sources of high energy 
photons are $\pi^0 \to \gamma \gamma$ and
$\eta \to \gamma \gamma$ decays. To reduce these backgrounds,
restrictions are imposed on the invariant mass of the candidate 
photon and any other photon ($\gamma^{\,\prime}$) in the event.
The candidate photon is rejected if 
120$\,$MeV/c$^2 < M(\gamma\gamma^{\,\prime}) < 145\,$MeV/c$^2$ and 
$E_{\gamma^{\,\prime}}\,>\,30\,$MeV or if
510$\,$MeV/c$^2 < M(\gamma\gamma^{\,\prime}) < 570\,$MeV/c$^2$ and
$E_{\gamma^{\,\prime}}\,>\,200\,$MeV.
To reduce the background from $\pi^0$ decays when the two daughter
photons form a single cluster in the calorimeter,
the ratio of the energy deposition in 3$\times$3 ECL cells compared to that in 
5$\times$5 cells around the maximum energy ECL cell
is required to exceed 95$\%$.

The $K^0_S$ candidates are formed from $\pi^+\pi^-$ combinations
that have an invariant mass within $\pm 10\,$MeV/c$^2$ of the nominal
$K^0_S$ mass ($\sim$ 3$\sigma$ in the $K^0_S$ mass resolution).
The two pions are required to have a common vertex 
that is displaced from the interaction point by more than 0.5\,cm in the
plane perpendicular to the beam direction.
The difference in $z$ coordinates for the tracks
constituting the secondary vertex must be less than 2\,cm.
The angle $\alpha$ between 
the $K^0_S$ flight direction and the measured $K^0_S$ momentum direction
is required to satisfy cos$\alpha\,>\,0.8$.
Opposite-sign $K$ mesons are combined to form
$\phi$ candidates; their invariant mass is required to be
within $\pm 10\,$MeV/c$^2$ ($\sim$ 3$\sigma$)
of the nominal value for the $\phi$ mass.

The $\phi K^- \gamma$ and $\phi K^0_S \gamma$ combinations are selected
to form $B^-$ and $\bar{B}^0$ candidates.
Two kinematic variables are used to extract 
the $B$ meson signal:
the energy difference $\Delta E\,=\,E^{*}_B-E^{*}_{\rm beam}$ and
the beam-constrained mass 
$M_{\rm bc}=\sqrt{(E^{*}_{\rm beam})^2\,-\,(p^{*}_B)^2}$,
where $E^{*}_B$ and $p^{*}_B$ are the CM energy and momentum
of the $B$ candidate and $E^{*}_{\rm beam}$ is the CM beam energy.
The events that satisfy loose requirements $M_{\rm bc} > 5.2$\,GeV/c$^2$ and 
$|\Delta E|\,<0.4$\,GeV are selected for further analysis.

With these selection criteria, the primary background source is continuum
$e^+e^- \rightarrow q \bar{q}$ production, where $q$ may be a
$u,\,d,\,s,$ or $c$ quark.
To separate spherical $B\bar{B}$ events from jet-like continuum events,
a Fisher discriminant is formed from six modified
Fox-Wolfram moments \cite{bellei,fox}.
Signal and background probability density functions (PDF) 
for the Fisher discriminant and the cosine of the $B$ flight direction
with respect to the $z$ axis (cos $\theta^*_B$) are obtained
from signal MC and sideband data.
The signal (background) PDFs are multiplied to form a
signal (background) likelihood ${\cal L}_S$ (${\cal L}_{BG}$).
The likelihood ratio 
\mbox{${\cal LR} = {\cal L}_S/({\cal L}_S + {\cal L}_{BG})$}
is required to be greater than 0.3.
This event topology
requirement retains 92$\%$ of the signal events while removing 55$\%$
of the continuum events.
Finally, the ratio of the second to the zeroth Fox-Wolfram moment,
calculated using all particles in the event, is required to be less than 0.5.

Signal MC studies of the $\Delta E$ distribution indicate that the width
is dominated by photon energy smearing, 
and the shape is expected to be asymmetric due to photon energy leakage.
The resolution in $M_{\rm bc}$ of 3.2\,MeV/c$^2$ is dominated 
by the beam energy
spread of KEKB and is slightly improved to 3.0\,MeV/c$^2$
by rescaling the photon energy so that 
\mbox{$\Delta E=E_{\phi}^{*}+\,E_K^{*}+\,E_{\gamma}^{*}-\,E^{*}_{\rm beam}=0$},
where $E_{\phi}^{*}$ and
$E_K^{*}$ are $\phi$ meson and $K$ meson energies in the CM frame.

The $M_{\rm bc}$ distribution for the interval
$-0.08\,$GeV $< \Delta E\,<0.05$\,GeV and $\Delta E$ distribution 
for the interval 5.27\,GeV/c$^2 < M_{\rm bc} < 5.29$\,GeV/c$^2$ 
for $\pkg$ decay mode are shown in Figs.~2(a) and 2(b).
To avoid systematic uncertainties
that could arise from the description of the $\Delta E$ distribution,
the signal yield is extracted from a fit of the 
$M_{\rm bc}$ distribution after applying the asymmetric cut
\mbox{$-0.08\,$GeV $< \Delta E\,<0.05$\,GeV}.
The $M_{\rm bc}$ distribution is fitted to the sum of a Gaussian function
and the so-called ARGUS background function \cite{argus}. The width of the
signal Gaussian is determined from MC, while the peak position is
fixed to 5.279$\,$GeV/c$^2$.
The background shape was studied using $M_{\rm bc}$ and $\phi$ mass
sidebands and found to be flat.
The fitted number of $B$ candidates is \mbox{$N=21.6 \pm 5.6$}.
As a crosscheck, 
the $\Delta E$ distribution was fitted with the Crystal Ball line shape 
function \cite{cb}
(the shape is fixed from MC)
to describe the signal and a linear function to describe the background. 
The signal yield is $N = 23.6 \pm 5.4$ events, which is consistent with
the $M_{\rm bc}$ fit result.
The $K^+K^-$ mass distribution with two combinations per event 
is shown in Fig.~2(c) for events in the 
$B$ signal region,
5.27$\,$GeV/c$^2 < M_{\rm bc} < 5.29$\,GeV/c$^2$ and 
$-0.08\,$GeV $< \Delta E\,<0.05$\,GeV.
The solid histogram in Fig.~2(c) shows events from 
the $B$ meson mass sideband, 5.2$\,$GeV/c$^2<M_{\rm bc}<5.26\,$GeV/c$^2$,
with a normalization obtained from the $M_{\rm bc}$ distribution fit.


\begin{figure}[t!]
\vspace{-0.5cm}
\begin{center}
\epsfig{file=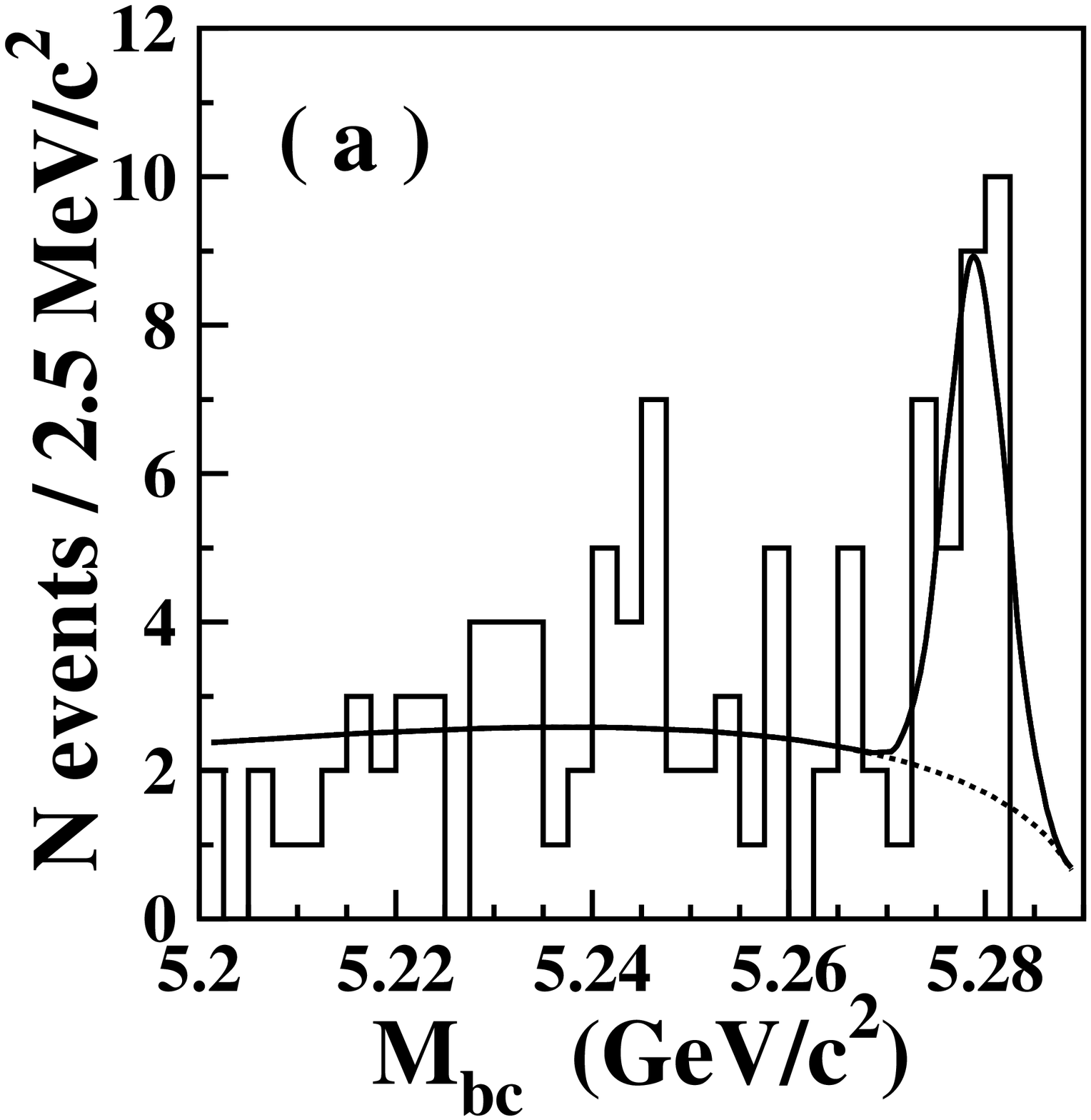,width=4.9cm,height=4.9cm}\epsfig{file=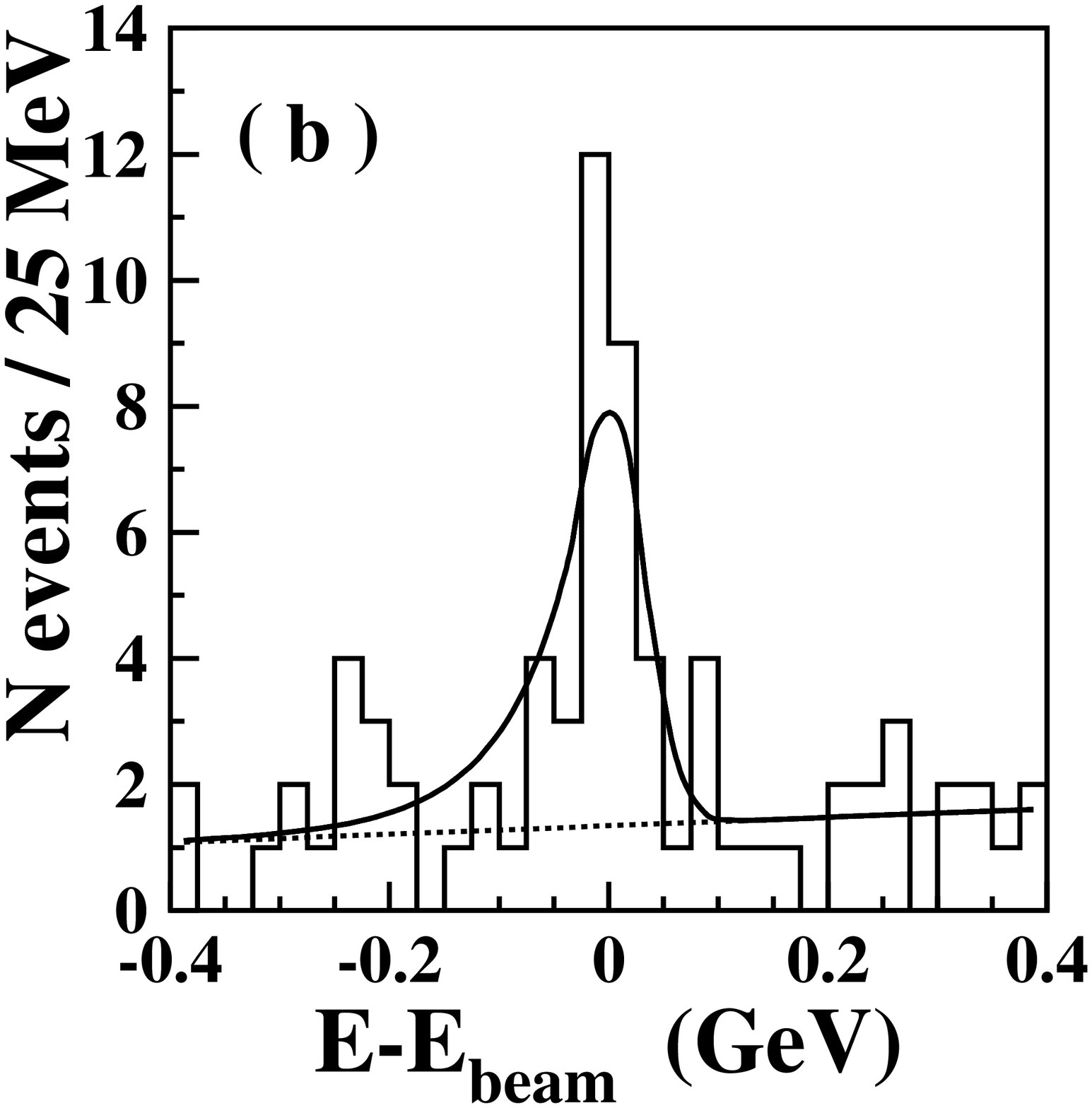,width=4.9cm,height=4.9cm}\epsfig{file=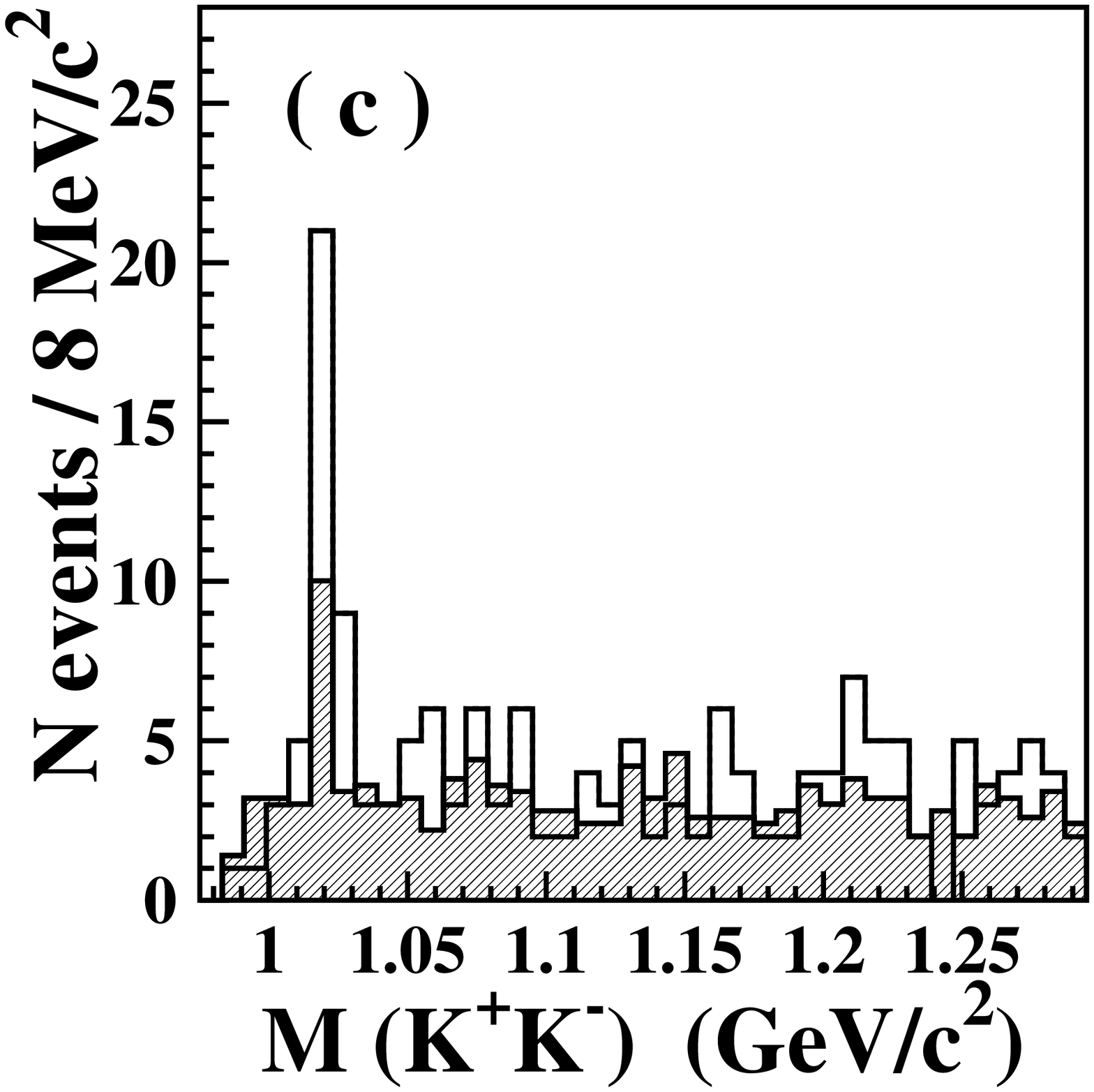,width=4.9cm,height=4.9cm}
\epsfig{file=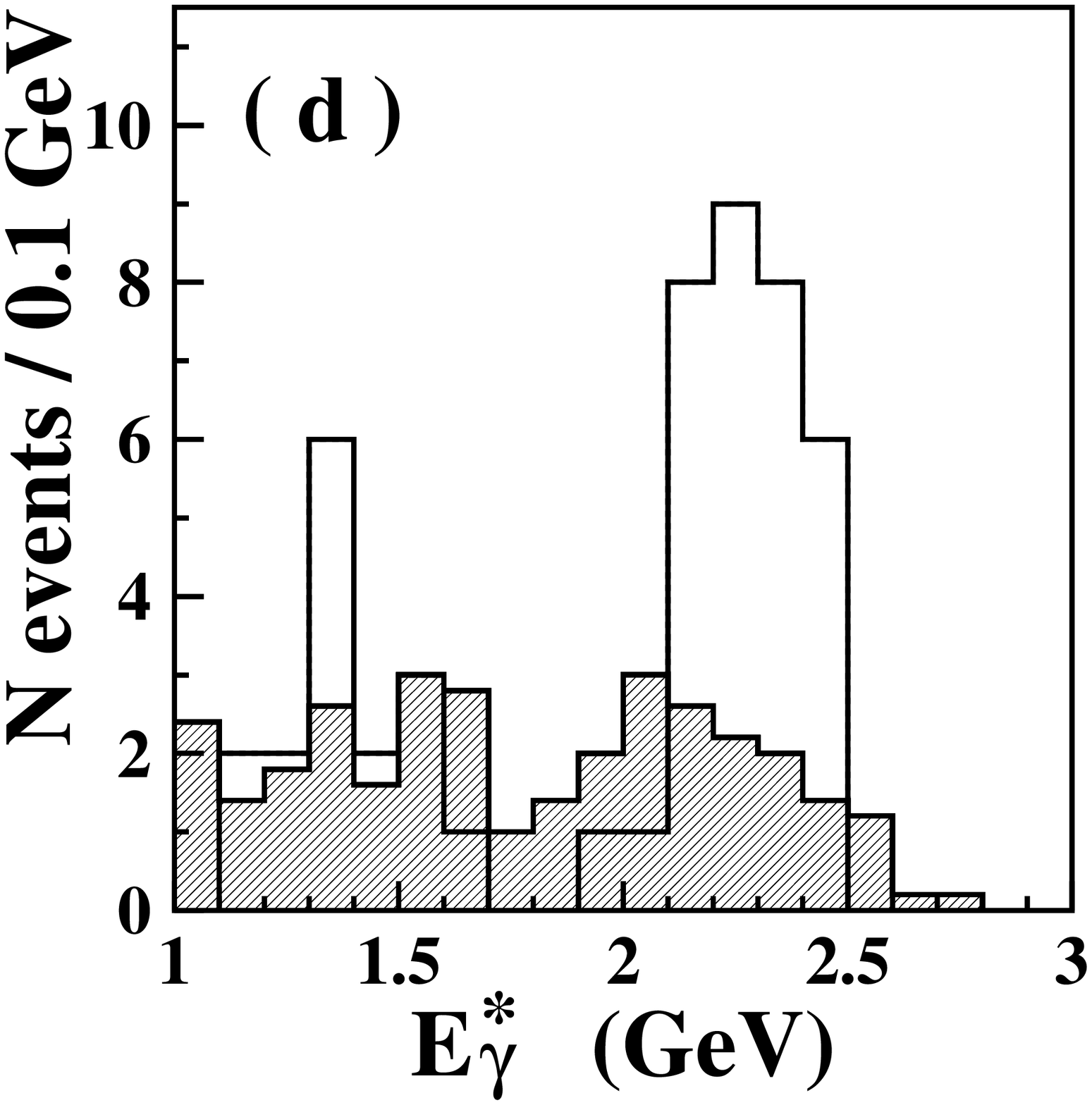,width=4.9cm,height=4.9cm}\epsfig{file=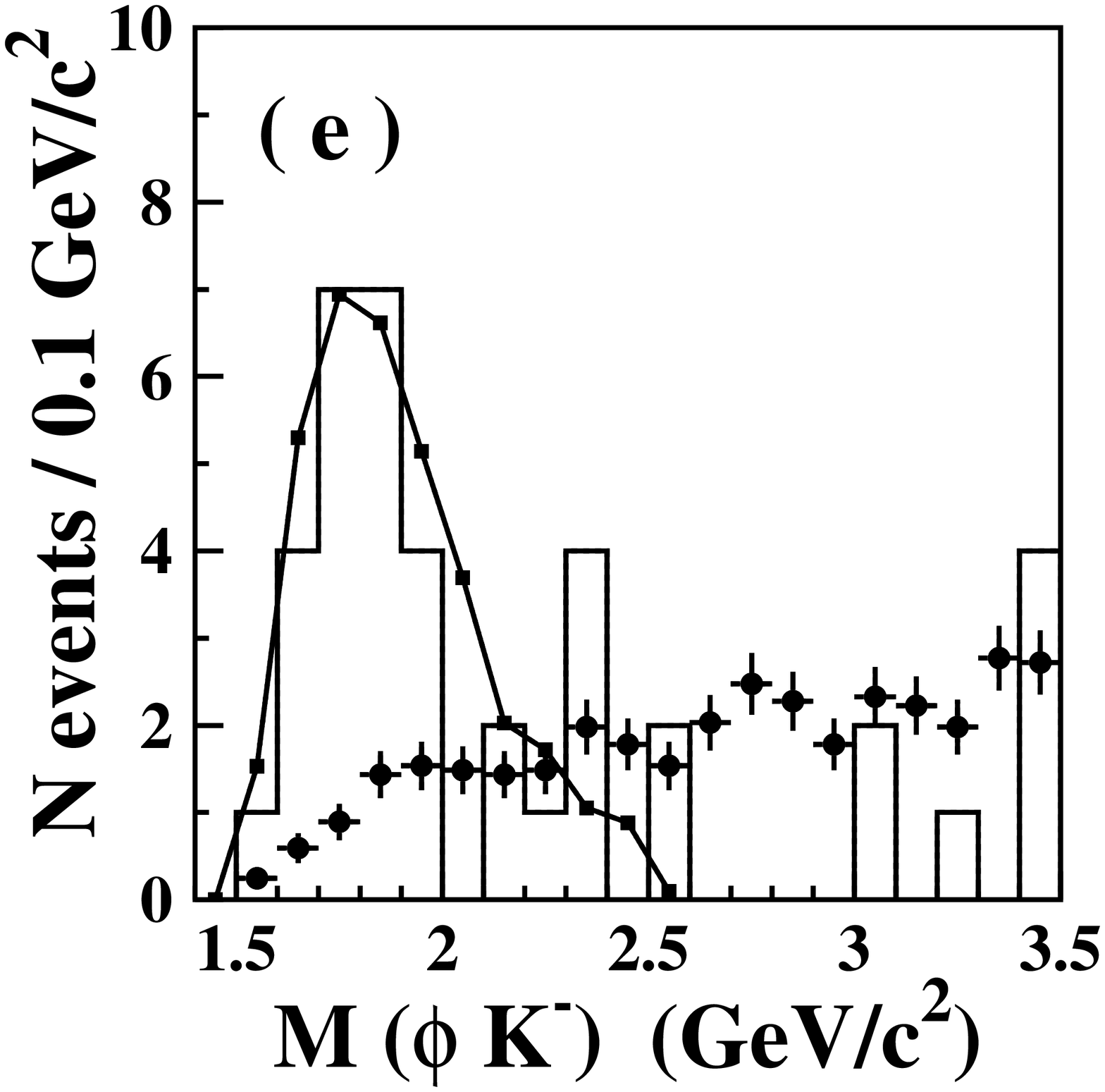,width=4.9cm,height=4.9cm}\epsfig{file=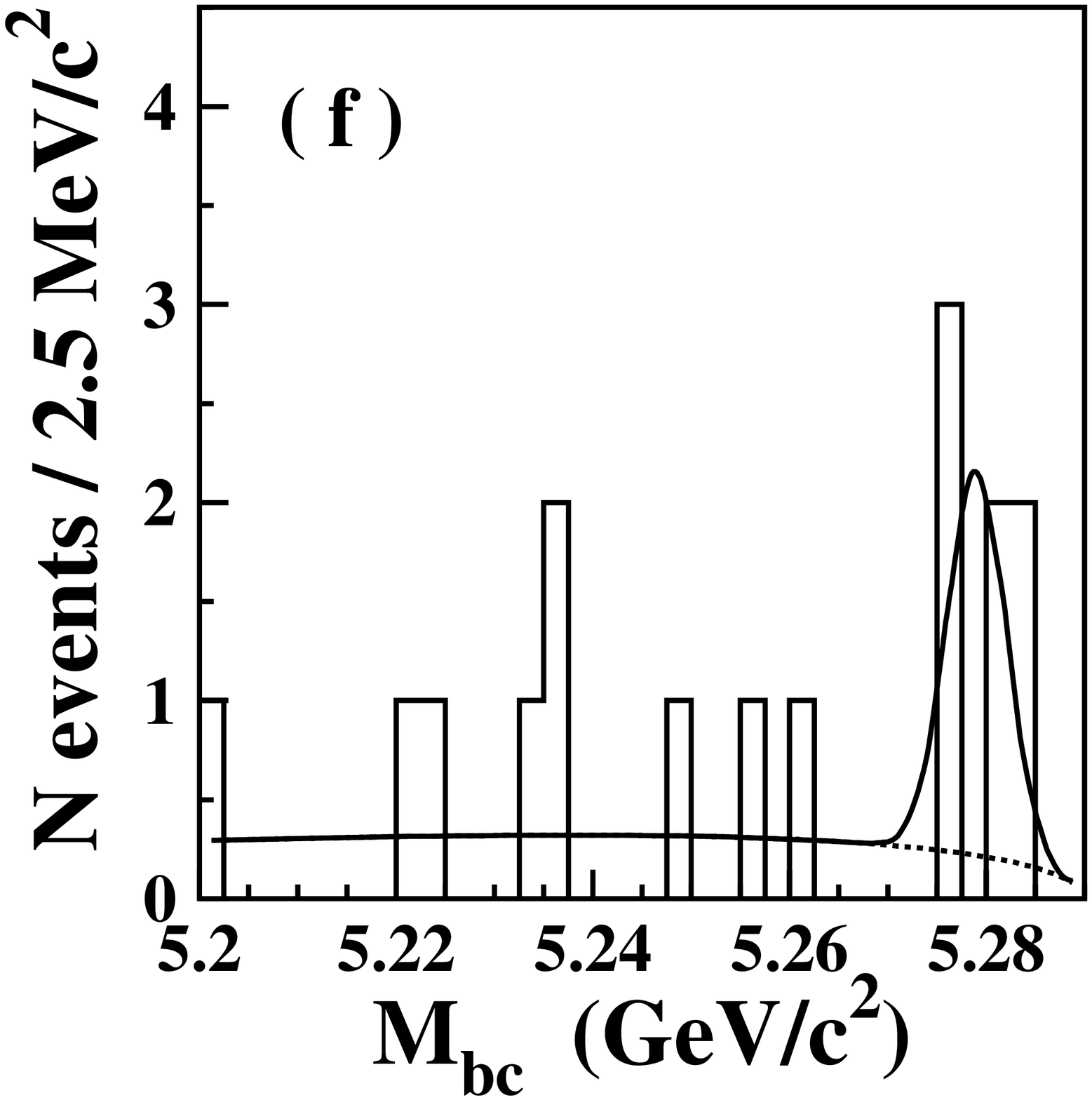,width=4.9cm,height=4.9cm}
\end{center}
\vspace{-0.4cm}
\caption{The (a) $M_{\rm bc}$, (b) $\Delta E$, (c) $M(K^+K^-)$, 
(d) $E^*_{\gamma}$ and (e) $M(\phi K^-)$ distributions 
for the $\pkg$ decay mode and the (f) $M_{\rm bc}$ distribution
for the $\pkog$ decay mode are shown by histograms.
The solid histograms are obtained with events from the $B$ mass
sideband.
The curves show the results of the fit described in the text. 
The measured $M(\phi K^-)$ distribution is compared to those 
obtained from the MC simulation,
with a three-body phase-space model (circles with error bars)
or adjusted to follow the data (squares connected by a line).}
\label{figall}
\end{figure}

The photon energy distribution for the $\pkg$ decay channel for events
within the $B$ signal region 
with an extended
photon energy interval 1.0\,GeV $< E^*_{\gamma} < 3.0\,$GeV
is shown in Fig.~2(d).
The solid histogram shows events from the $B$ mass 
sideband,
which are used to describe the background.
The MC determined $B$ meson reconstruction efficiency is nearly 
constant for photon energies between 1.5\,GeV and 2.7\,GeV.
Figure 2(d) demonstrates that the signal events are concentrated in
the range 2.1$\,$GeV $< E^*_{\gamma} < 2.5\,$GeV,
and events outside that range are consistent with
background, estimated from the $B$ mass sideband events.
The bulk of inclusive $B \rightarrow X_s \gamma$ events were also observed 
in the $E^*_{\gamma} > 2.0\,$GeV range \cite{bellei,babari,cleoi}. 
The maximum photon energy is kinematically constrained by the mass of the 
hadronic system and is therefore slightly lower in $\pkg$ decays 
compared to $B \rightarrow X_s \gamma$.

To search for a possible contribution from kaonic resonances decaying to 
$\phi K^-$, the $\phi K^-$ invariant mass distribution
is also shown in Fig.~2(e). 
This distribution is strongly correlated with 
the photon energy distribution.
Background was not subtracted in Fig.~2(e) because of
the small number of signal and background events.
Events mostly populate the low-mass range; however, the small data sample 
precludes any definite conclusions. It is clear that the observed 
$\phi K^-$ mass distribution differs significantly from that in the 
three-body phase-space decay (also shown in Fig.~2(e)). 
To provide the correct branching fraction 
measurement, the MC data sample was adjusted to follow the measured 
$\phi K^-$ invariant mass distribution.

The $M_{\rm bc}$ distribution for the $\pkog$ decay channel
is shown in Fig.~2(f). The figure is obtained in a similar manner 
to those for $\pkg$ decay.
In the fit of the $M_{\rm bc}$ distribution the peak position 
is fixed to the value 5.279$\,$GeV/c$^2$
and the background shape (defined by the slope parameter
of the ARGUS function) is constrained to be flat.

The signal yields, efficiencies, branching fractions and significances
obtained for the $\pkg$ and $\pkog$ decay channels 
for the photon energy range 2.0\,GeV $< E^*_{\gamma} < 2.7\,$GeV
are given in Table 1.
The signal in the decay channel $\pkg$ has a
5.5 $\sigma$ significance, whereas that for $\pkog$ is
3.3 $\sigma$ and a 90$\%$ confidence level upper limit,
which is calculated assuming a Gaussian distribution of the statistical
error and an additional one unit of the systematic error contribution,
is also given.
The statistical significance is defined as 
$\sqrt{-2\,\ln({\cal L}_0/{\cal L}_{max})}$,
where ${\cal L}_{max}$ and ${\cal L}_0$ are likelihood 
values at the best-fit signal yield
and the signal yield fixed to zero. 
The branching fractions for the decays of the intermediate $\phi$ and
$\bar{K}^0$ states are taken from Ref.~\cite{pdg} and  
are not included in the efficiencies quoted in Table 1.
No event with more than one $B$ candidate is 
found in the data; 
the double counting probability for the $\pkg$ channel is
estimated by MC to be $(1.2 \pm 0.3)\%$ and is
accounted for in the efficiency calculation.
An equal production rate for the neutral and charged $B$ mesons
is assumed. 

\begin{table}[h!]
\caption{The signal yields obtained from the $M_{\rm bc}$ fit,
efficiencies, branching fractions, upper limit
and significances for the $\pkg$ and $\pkog$ decay modes.}
\vspace{0.2cm}
\label{tab:bfr}
\begin{tabular}
{@{\hspace{0.5cm}}l@{\hspace{0.5cm}}||@{\hspace{0.5cm}}c@{\hspace{0.5cm}}||@{\hspace{0.5cm}}c@{\hspace{0.5cm}}||@{\hspace{0.5cm}}c@{\hspace{0.5cm}}||@{\hspace{0.5cm}}c@{\hspace{0.5cm}}}
\hline
\hline
Decay mode & Yield & Efficiency & Branching fraction & Significance \\
      &  & ($\%$) & ($10^{-6}$) & ($\sigma$) \\ \hline \hline
  $\pkg$ & 21.6 $\pm$ 5.6 & 12.9 & 3.4 $\pm$ 0.9 $\pm$ 0.4 & 5.5 \\
 $\pkog$ & 5.8 $\pm$ 3.0 & 7.9 & 4.6 $\pm$ 2.4 $\pm$ 0.6 & 3.3 \\
         &               &     & $<8.3\ (90\%\ CL)$ & \\ \hline\hline
\end{tabular}
\vspace{-0.1cm}
\end{table}

In addition to the dominant continuum background, various $B\bar{B}$
background sources were studied. 
The only significant background found,
with a contribution exceeding $1\%$ of the signal level, arises from 
non-resonant $B^- \to K^- K^- K^+ \gamma$ decays.
This contribution is estimated to be $(4 \pm 4)\,\%$
using events from the high mass $\phi$ meson sideband 
1.05\,GeV/c$^2<\,M_{K^+K^-}\,<1.25\,$GeV/c$^2$.
The shape of this background as a function of $M_{K^+K^-}$ 
is taken from the MC simulation, assuming non-resonant decay
$K^{*-}(1770) \to K^- K^- K^+$.
The $\pkg$ branching fraction is corrected for this contribution, and 
the uncertainty is included in the systematic error.

The major sources contributing to the systematic
uncertainty of the branching fraction measurements are:
the photon reconstruction efficiency and energy scale
(4\%),
the reconstruction efficiency of charged tracks (1\% per track),
the charged kaon particle identification (1\% per particle),
the neutral kaon reconstruction efficiency (3\%), 
uncertainties in the non-resonant background under the $\phi$ signal
(4\%),
uncertainties in the angular distributions applied in the 
MC simulation of the hadronic states (5\%),
uncertainties in the MC simulation of $\pi^0$ and $\eta$ rejection criteria
(2\%),
uncertainty in the efficiency of the topological likelihood cut (2\%),
uncertainties of the background and $B$ signal shape description
in the fit (5\%),
and uncertainty in the determination of the number of $B\bar{B}$ pairs
($<$ 1\%). The systematic uncertainties described above are 
added in quadrature to obtain the final systematic errors.

In conclusion, the $\apkg$ decay modes were studied for the first time. 
A branching fraction for the $\pkg$ decay mode and an upper limit for
the $\pkog$ decay mode were obtained.
The measured $\pkg$ decay branching fraction is roughly an order of
magnitude smaller than the $B \rightarrow K^{*} \gamma$
branching fractions.
This factor can be 
attributed to the creation of the additional $s\bar{s}$ pair.
Signal events are concentrated in the photon energy
range \mbox{2.1$\,$GeV$\,<\,E^*_{\gamma}\,<\,2.5\,$GeV},
which is similar to the photon energy range for 
two-body radiative decays.

We wish to thank the KEKB accelerator group for the excellent
operation of the KEKB accelerator.
We acknowledge support from the Ministry of Education,
Culture, Sports, Science, and Technology of Japan
and the Japan Society for the Promotion of Science;
the Australian Research Council
and the Australian Department of Education, Science and Training;
the National Science Foundation of China under contract No.~10175071;
the Department of Science and Technology of India;
the BK21 program of the Ministry of Education of Korea
and the CHEP SRC program of the Korea Science and Engineering Foundation;
the Polish State Committee for Scientific Research
under contract No.~2P03B 01324;
the Ministry of Science and Technology of the Russian Federation;
the Ministry of Education, Science and Sport of the Republic of Slovenia;
the National Science Council and the Ministry of Education of Taiwan;
and the U.S.\ Department of Energy.

\newpage

\end{document}